\documentstyle[12pt,aaspp4,flushrt]{article}

\newcommand{\etal}{{\rm et al.~}}
\newcommand{\Mpc}{$h^{-1}$~{\rm Mpc}}
\newcommand{\hmpc}{$h$~{\rm Mpc$^{-1}$}}

\begin{document}

\title{Steps toward the power spectrum of matter.
II. The biasing correction with $\sigma_8$ normalization}

\author{J. Einasto \altaffilmark{1},
{M. Einasto} \altaffilmark{1},
{E. Tago} \altaffilmark{1}, 
{V. M\"uller}\altaffilmark{2},
{A. Knebe}\altaffilmark{2}, 
{R. Cen}\altaffilmark{3}, 
{A. A. Starobinsky} \altaffilmark{4}, and
{F. Atrio-Barandela }\altaffilmark{5}
}

\altaffiltext{1}{Tartu Observatory, EE-2444 T\~oravere, Estonia}
\altaffiltext{2}{Astrophysical Institute Potsdam, An der 
Sternwarte 16,  D-14482 Potsdam, Germany} 
\altaffiltext{3}{Department of Astrophysical Sciences, Princeton University,
Princeton, NJ 08544, USA }
\altaffiltext{4}{Landau Institute for Theoretical Physics, Moscow 117334, 
Russia} 
\altaffiltext{5}{F{\'\i}sica Te\'orica, Universidad de Salamanca, 
37008 Spain }

\begin{abstract}

We suggest a new method to determine the bias parameter of galaxies
relative to matter. The method is based on the assumption that gravity
is the dominating force which determines the formation of the
structure in the Universe. Due to gravitational instability matter
flows out of under-dense towards over-dense regions. To form a
galaxy, the density of matter within a certain radius must exceed a
critical value (Press-Schechter limit), thus galaxy formation
is a threshold process. In low-density environments (voids) galaxies
do not form and matter remains in primordial form.  We estimate the
value of the threshold density which divides the matter into two
populations, a low-density population in voids and a clustered
population in high-density regions.  We investigate the influence of
the presence of these two populations to the power spectrum of matter
and galaxies. We find that the power spectrum of clustered particles
(galaxies) is similar to the power spectrum of matter.  We show that
the fraction of total matter in the clustered population determines
the difference between amplitudes of fluctuations of matter and
galaxies, i.e. the bias factor.

To determine the fraction of matter in voids and clustered population
we perform numerical simulations. The fraction of matter in galaxies
at the present epoch is found using a calibration through the
$\sigma_8$ parameter.  We find $\sigma_8=0.89 \pm 0.09$ for galaxies,
$\sigma_8 = 0.68 \pm 0.09$ for matter, and $b_{gal}=1.3 \pm 0.13$ --
the biasing factor of the clustered matter (galaxies) relative to all
matter.
 
\end{abstract}



\section{Introduction}

The relative distribution of matter and light in the Universe is an
unresolved problem of fundamental importance in cosmology. Direct
observational data on galaxies give information on the distribution of
light; in contrast, most theoretical models simulate only the
distribution of matter. That these two distributions may be different
on galactic scales was evident since the discovery of dark halos
around galaxies (Einasto, Kaasik \& Saar 1974a, Ostriker, Peebles \&
Yahil 1974).  Several years later it became clear that differences
between the distribution of light and matter exist on large scales as
well.  J\~oeveer \& Einasto (1978, hereafter JE78) demonstrated that
galaxies and clusters of galaxies are distributed in filaments, and
the space between them is practically void of visible matter, whereas
in numerical simulations of structure formation (Zeldovich 1978)
low-density regions are not completely empty of matter.  This
difference between the distribution of galaxies and dark matter (DM)
was quantified by Zeldovich, Einasto \& Shandarin (1982): in numerical
simulations there exists a population of almost isolated particles in
voids which has no counterpart in the observed distribution of
galaxies.  Einasto, J\~oeveer \& Saar (1980, hereafter EJS80) showed
that this difference can be explained if the evolution of the Universe
is primarily due to gravity.  As demonstrated by Zeldovich (1970,
hereafter Z70), the gravitational instability enhances the density
contrast: matter flows out from low-density regions toward
high-density ones until it collapses to form galaxies and systems of
galaxies. This process is slow and gravity is not able to evacuate
voids completely -- there must exist some primeval matter in voids.

Bahcall \& Soneira (1983) and Klypin \& Kopylov (1983) demonstrated
that the correlation function of clusters of galaxies has an amplitude
larger than that of galaxies, and Kaiser (1984) explained this using
the theory of high peaks in a Gaussian density field, introducing the
term ``biasing'' to describe this fact. A similar relation holds for
power spectra of clusters and galaxies; and we define the bias
parameter $b_c$ through the power spectra of all matter, $P_m(k)$, and
that of the clustered matter, $P_c(k)$,
$$
P_c(k) = b_c^2(k) P_m(k),
\eqno(1)
$$
where $k$ is the wavenumber in units of $h$~Mpc$^{-1}$, and the Hubble
constant is expressed as $H_0 = 100~h$ km~s$^{-1}$~Mpc$^{-1}$.
According to this definition, the biasing parameter is a function of
wavenumber $k$.  The power spectrum is calculated by integrating the
density contrast over the whole space under study, thus the biasing
parameter is a mean averaged over the space.

As the bias factor of galaxies relative to matter was not known it
was considered as a free parameter suitable to bring models of
structure formation into agreement with observations of density
fluctuations of galaxies. As an example we refer to the pioneering
study of the standard CDM model by Davis \etal (1985). Here a large
biasing factor $b=2.5$ was applied to make the model agree with
observations.  Actually the bias factor is a fundamental parameter
characterizing the distribution of matter and galaxies, and must be
determined from data.

In this paper we concentrate on the problem of how the presence of
primordial matter in voids affects the power spectrum.  The idea we
shall use of how to estimate the bias factor of clustered matter
relative to all matter was suggested by Gramann and Einasto (1992,
hereafter GE92). They assumed that the structure evolution of the
Universe is primarily due to gravity. The primordial matter contracts
and eventually forms galaxies only in case when its density is high
enough, in other words, the formation of galaxies is essentially a
threshold phenomenon (EJS80, Einasto \& Saar 1986). GE92 demonstrated
that the power spectra of matter and clustered matter (galaxies) are
similar in shape, and that the relative amplitude of the power
spectrum of galaxies (clustered population) depends on the fraction of
matter in the clustered population.  We consider as ``clustered
matter'' all matter associated with galaxies, including dark halos of
galaxies, and clusters of galaxies.  Einasto \etal (1994, hereafter
E94) studied the evacuation of voids and estimated the biasing
parameter of clustered matter relative to all matter.

Here we shall investigate the relation between the power spectra of
clustered matter and all matter in more detail, and derive a new
estimate of the respective bias factor.  The paper is organized as
follows. In Section 2 we consider the biasing as a physical
phenomenon and compare our approach with other biasing studies.
Thereafter we investigate the influence of the void matter on the
power spectra of galaxies and matter, using numerical simulations. One
problem with numerical simulations is the identification of the present 
epoch. We do this in Section 3 using the $\sigma_8$ normalization. In
Section 4 we discuss our results.  Section 5 gives main conclusions
of the study.

\section{Power spectra of galaxies and matter}

\subsection{Physical biasing}

We shall assume here that the structure evolution of the Universe is
basically due to gravity, that initial density fluctuations are
Gaussian and adiabatic (i.e. velocities of particles are in agreement
with the density field).  Due to gravitational instability, the
evolution of matter in under- and over-dense regions is different.
Gravity attracts matter toward high-density regions, thus particles
flow away from low-density regions, and density in high-density
regions increases until contracting objects collapse.  As shown by
Z70, the collapse occurs along caustics.  Initially it was assumed
that caustics are two-dimensional pancake-like structures. However,
direct observational data (JE78, EJS80, Einasto, Klypin \& Shandarin
1983) show that basic structural elements of the Universe are strings
and chains of galaxies and clusters.  Numerical simulations of
structure evolution have confirmed the formation of essentially
one-dimensional structures if the effective power index of the
spectrum on galactic scales is negative (Melott \etal 1983, Einasto
\etal 1986, Einasto \& Saar 1987, Gramann 1988, Melott \& Shandarin
1993, Pauls \& Melott 1995, Brodbeck \etal 1998).  Bond, Kofman \&
Pogosyan (1996) demonstrated analytically that under very general
assumptions the gravitational evolution leads to a network (web) of
high-density filaments and low-density regions outside of the web.
General physical considerations suggest that in high-density regions
the gas is cooling which creates favorable conditions for star
formation (Rees \& Ostriker (1977), Silk (1977)).  Thus the
gravitational character of the evolution leads to the formation of a
filamentary web of galaxies and clusters of galaxies.

According to Press \& Schechter (1974), the contraction occurs if the
linear over-density exceeds a factor of 1.68 in a sphere of radius $r$
which determines the mass of the system or the galaxy.  In a
low-density environment the matter cannot contract and remains
primordial.  Hydrodynamical simulations of galaxy formation by Cen and
Ostriker (1992, 1998), Katz, Hernquist \& Weinberg (1992), Katz,
Weinberg \& Hernquist (1996), Weinberg, Katz \& Hernquist (1997) have
confirmed that the galaxy formation is ineffective in low-density
environment. Gas cooling, as required for star formation, occurs only
in high-density regions -- near the centers of contracting clumps of
primordial matter.

These considerations lead us to the conclusion, that within the
gravitational instability picture the formation of galaxies is a
density threshold phenomenon (EJS80, Einasto \& Saar 1986). The central
problems are, how to find the threshold density which divides the
primordial matter in low-density regions and the clustered matter in
high-density regions, and how the division of matter into under-dense
and over-dense populations influences the power spectra of galaxies
and systems of galaxies.

\subsection{Density field of galaxies and matter}

The true density field of the Universe is a continuous function of
coordinates since the dominating population is dark matter (DM) which
consists of particles of small mass. Here we accept the current
paradigm that DM (or at least, most of it) is non-baryonic.  The
dynamical evolution of the Universe can be simulated using N-body
calculations. As the number of particles in simulations is limited, it
is impossible to simulate the motion of all DM particles; in practice
the mass of particles in simulations is usually a fraction of the mass
of a typical galaxy.  Thus, in order to find the true density field of
DM, the distribution of a discrete set of particles is to be smoothed.
Galaxies also are discrete objects, the density field of galaxies can
be calculated by smoothing, too.  Here the smoothing length is of
prime importance.

Galaxies form mostly in small groups which are collected from
primordial matter in a comoving volume of a Megaparsec scale.  Thus,
to start such a collapse of primordial matter, regions are needed with
at least the mean matter density when smoothed on Megaparsec
scales. Presently DM forms halos around galaxies in clusters and
groups, and the characteristic scale of halos is the semi-minor radius
of these systems, also $\approx 1$~\Mpc\ (Einasto \etal 1984). In the
present paper our goal is to find {\em the true density field of DM}
as accurately as possible.  We conclude that the density field has to
be found from positions of galaxies or simulation particles with a
$\sim 1$~\Mpc\ smoothing length.  We call densities calculated with a
small (one Mpc scale) smoothing parameter as {\em local} ones, in
contrast to {\em global} densities which are found using a large (ten
Mpc scale) smoothing length.

\subsection{Comparison with conventional biasing studies}

Our approach to the biasing phenomenon differs from the approach of
most other investigators. Commonly the biasing parameter is defined as
the ratio of the density contrast of galaxies and matter at location
${\bf x}$,
$$
\delta_{gal}({\bf x})= b \delta_m({\bf x}).
\eqno(2)
$$
As there are no galaxies in voids, we expect $b=0$ there.  If galaxies
trace the matter in high-density regions, then in these regions $b=1$.
In order to apply this formula and to find a mean value of the biasing
parameter, the density field is conventionally smoothed with a rather
large smoothing length ($\geq 8$~\Mpc, see Dekel \& Lahav (1998)),
Blanton \etal (1998) and references therein for recent studies).
Excessive smoothing mixes unclustered DM in voids and clustered DM in
high-density regions which makes the simple biasing phenomenon rather
complicated.

Our experience with density smoothing has shown that large smoothing
is useful if one wants to locate large high-density regions, such as
superclusters of galaxies, see Figure~1b of Lindner \etal (1995) and
Figure~2 of Frisch \etal (1995).  In this case the true filamentary
nature of galaxy systems is completely lost, and we can only see the
distribution of large over- and under-dense regions.  Visualizations
of high-resolution simulations of structure evolution show that
high-density regions of gas and dark matter form an almost coinciding and
very thin filamentary web (for a recent study see Brodbeck \etal
1998), confirming older results obtained with lower resolution.  Thus,
if we are interested in the true density field of dark matter we have
to use a small smoothing length.

\subsection{The influence of a homogeneous population}

The population of clustered particles is obtained from the population
of all particles by exclusion of void particles. Now we shall analyze
how the exclusion of void particles influences the power spectrum. 
The power spectrum is defined through the density contrast 
$$
\delta({\bf x})={\varrho({\bf x}) - \bar\varrho \over \bar\varrho};
\eqno(3)
$$
here $\varrho({\bf x})$ is the density at location ${\bf x}$, and
$\bar\varrho$ is the mean density.

Consider an idealized density field, which consists of a fluctuating
clustered component and a background of constant density, so that 
$$
\varrho_m({\bf x}) = \varrho_c({\bf x}) + \varrho_s({\bf x});
\eqno(4)
$$
here subscripts $m$, $c$, and $s$ are for all matter, and its 
clustered and smooth components, respectively.
The density contrast of the matter is 
$$
\delta_{m}= {\varrho_{m} - \bar\varrho_{m} \over \bar\varrho_{m}};
$$
or, applying (4),   
$$
\delta_{m}= {\varrho_c + \varrho_s -(\bar\varrho_c +
\bar\varrho_s)
  \over \bar\varrho_c + \bar\varrho_s }.
$$
Since  $\varrho_s=\bar\varrho_s$,
$$
\delta_{m}= {\varrho_c -\bar\varrho_c \over \bar\varrho_m} =
\delta_{c} {\bar\varrho_c \over \bar\varrho_m}.
$$
In the last equation $\bar\varrho_c/\bar\varrho_{m}$ is the
fraction of matter in the clustered population, $F_c$; and we get
$$
\delta_m = \delta_c F_c.
\eqno(4')
$$

A similar formula holds for the density contrast in Fourier space, and we
obtain the relation between power spectra of matter and the clustered
population 
$$
P_m(k)= \langle|\delta_m(k)|^{2}\rangle =  F_c^2 P_c(k),
\eqno(5)
$$
where $\delta_m(k)$ is the Fourier component of the matter density
contrast for a wavenumber $k$; and the averaging is over the whole
space under study.  We see that for this ideal case the power spectra
of matter and that the clustered population are related by an equation
similar to (1). Hence we get for the bias factor
$$
b_c={1 \over F_c}.
\eqno(6)
$$

Equations (5) and (6) were derived by GE92. These equations show that
the subtraction of a homogeneous population from the whole matter
population increases the amplitude of the spectrum of the remaining
clustered population.  In this approximation biasing is linear and
does not depend on scale.  These equations have a simple
interpretation.  The power spectrum describes the square of the
amplitude of the density contrast, i.e. the amplitude of density
perturbations with respect to the mean density.  If we subtract from the
density field a constant density background but otherwise preserve
density fluctuations, then amplitudes of {\em absolute} density
fluctuations remain the same, but amplitudes of {\em relative}
fluctuations with respect to the mean density increase by a factor
which is determined by the ratio of mean densities, i.e. by the fraction of
matter in the new density field with respect to the previous one.

The density of the real void population is not constant, neither is it
appropriate to attribute part of the dark matter particles in
high-density regions to the smooth background, i.e. to the void
population, which does not penetrate high-density regions.  Thus we
have to ask: How are power spectra of matter and the clustered
population related for a more realistic distribution of matter?

\subsection{The evolution of low- and high-density regions
in simulations}

To answer this question we performed numerical simulations of the
evolution of matter. As the above formulae are identical in the 2-D and
3-D cases we used a 2-D simulation to obtain a better resolution. We
used a particle-mesh (PM) algorithm with $512^{2}$ particles and
cells, and a double-power law initial power spectrum, (eqn. (3) in
Einasto \etal (1999a), Paper I) -- a simple approximation of the
observed spectra of galaxies and clusters of galaxies with a sharp
maximum (Frisch \etal 1995), as seen in Figure~1 of Paper I.  The
power index on large scales (Harrison-Zeldovich region) was taken to
be $n=2$, and $m=-1$ on small scales; in the 3-D case these indices
correspond to $n=1$ and $m=-2$ on large and small scales,
respectively.  The turnover at $k_0$ was $L/4$, where $L=512$~\Mpc\ is
the box size. The corresponding linear scale is $l_{max} =
128$~\Mpc. The present epoch was identified using an rms density
dispersion of $\sigma_1=4$ on a scale of 1~\Mpc, which corresponds 
approximately to a variance of $\sigma_{8}=0.9$ on a scale of 8~\Mpc.

We find the density field using a top-hat smoothing on 1~\Mpc\ scale
(cell size), and, by linear interpolation in both coordinates, we
attribute a density value to each particle. Densities are expressed in
units of the mean density.  We assume that particles with different
density labels can be used to represent void particles and galaxies
(with their halos) of different morphological type and environment.
We shall discuss this assumption and the relation between clustered
particles and galaxies in more detail in the next subsection.

In accordance with arguments discussed above we call all particles
with low density values ($\varrho < \varrho_{0}$) {\it void
particles}; all others are called {\it clustered particles}.
Particles with high density values ($\varrho \geq \varrho_{cl}$) are
associated with clusters or groups, and particles with intermediate
density values ($\varrho_{0} \leq \varrho < \varrho_{cl}$) with field
galaxies.  In the real Universe field galaxies are located around
clusters and form filaments between clusters and groups.  Here
$\varrho_{0}$ and $\varrho_{cl}$ are the threshold densities that
divide void particles from clustered ones, and particles associated
with clusters from particles bound to field galaxies.

Of course, in the real Universe the threshold between void and
clustered particles is not sharp.  If a clump of primordial matter is
small enough, then it can contract and form a dwarf galaxy, even if
the density, smoothed on 1 \Mpc\ level, is smaller than $\varrho_0$.
Similarly, if a clump is large, it can remain in primordial form if the
density, smoothed on 1 \Mpc\ level, is greater than $\varrho_0$. Such
local irregularities make the threshold fuzzy.  What matters is the
mean value of the threshold.  We investigate the influence of the
fuzziness of the threshold density to the power spectrum of galaxies
below. 

\begin{figure}[ht]
\vspace*{14.5cm} \figcaption{ The distribution of simulated and real
galaxies in a box of side-length 90~\Mpc.  Panel (a) gives particles
in voids ($\varrho <1$); panel (b) shows the distribution of simulated
galaxies in high-density regions: galaxies in the density interval $5
\leq \varrho < 20$ are plotted as black dots, galaxies with $\varrho
\geq 20$ as filled (red) regions; panel (c) shows field galaxies in
the density interval $1 \leq \varrho <1.5$ (open blue circles), and
$1.5 \leq \varrho < 5$ (dots).  Densities are expressed in units of
the mean density of the Universe.  Panel (d) shows the distribution of
galaxies in supergalactic coordinates in a sheet $0 \leq X < 10$~\Mpc,
horizontal and vertical axes are supergalactic $Y$ and $Z$,
respectively; bright galaxies ($M_B \leq -20.3$) are plotted as red
dots, galaxies $-20.3 < M_B \leq -19.7$ as black dots, galaxies $-19.7
< M_B \leq -18.8$ as open blue circles, galaxies $-18.8 < M_B \leq
-18.0$ as green circles (absolute magnitudes correspond to Hubble
parameter $h=1$).  High-density regions are the Local, the Coma and
the Hercules superclusters in lower left, lower right and upper right
corners, respectively. The long chain of galaxies between Coma and
Hercules superclusters is called the Great Wall, actually it is a
filament.  }  
\includegraphics{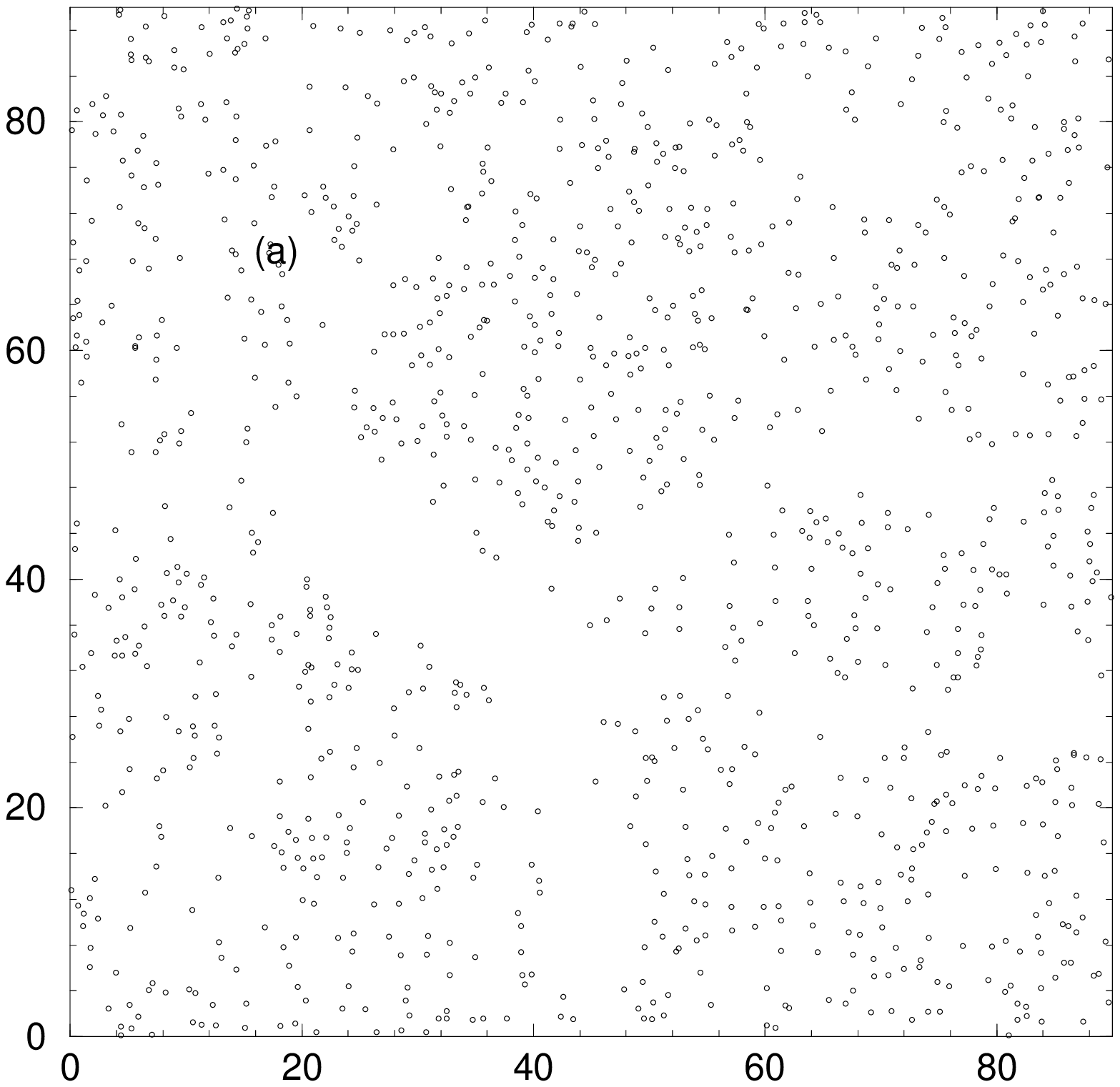} 
\includegraphics{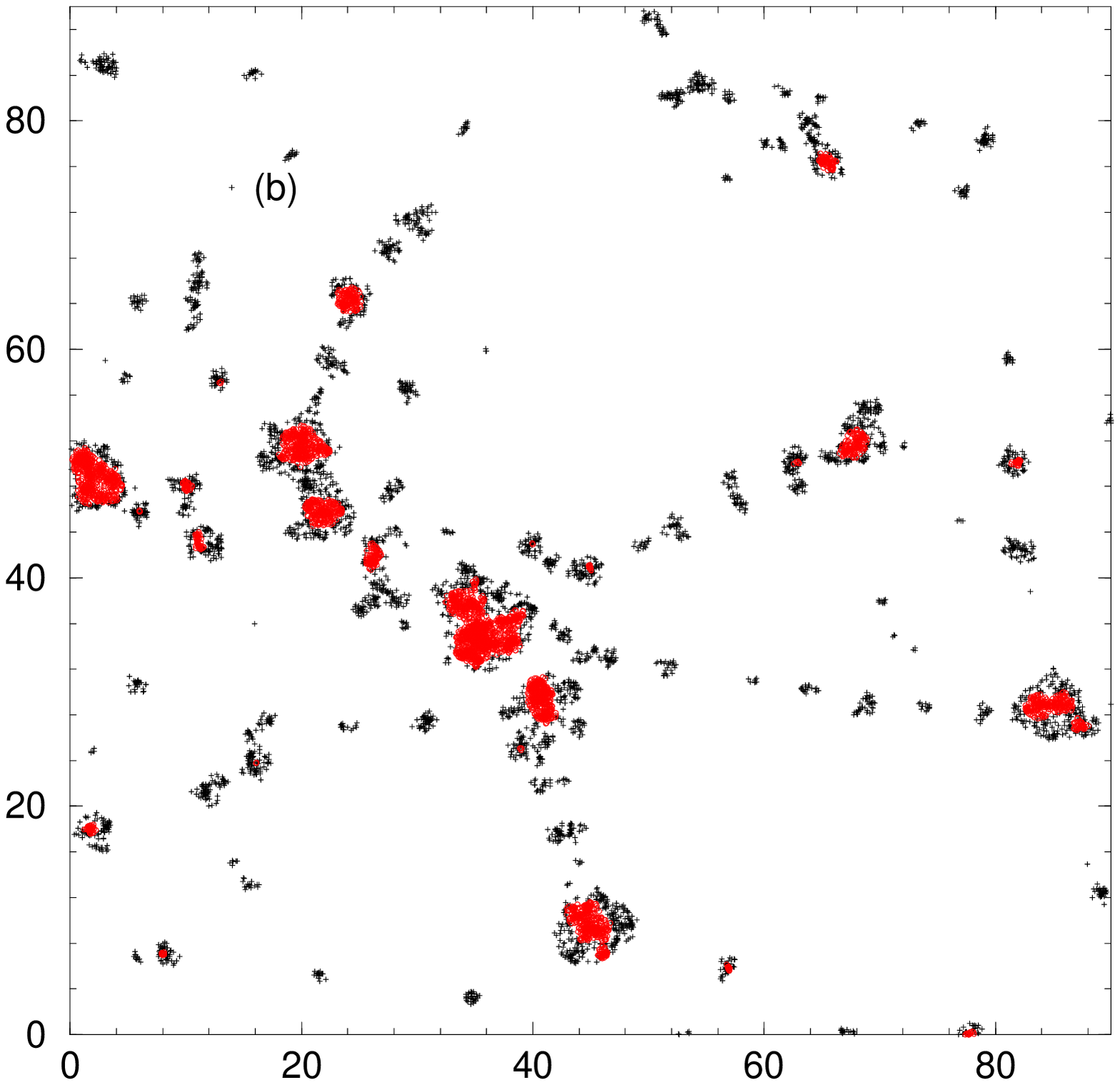} 
\includegraphics{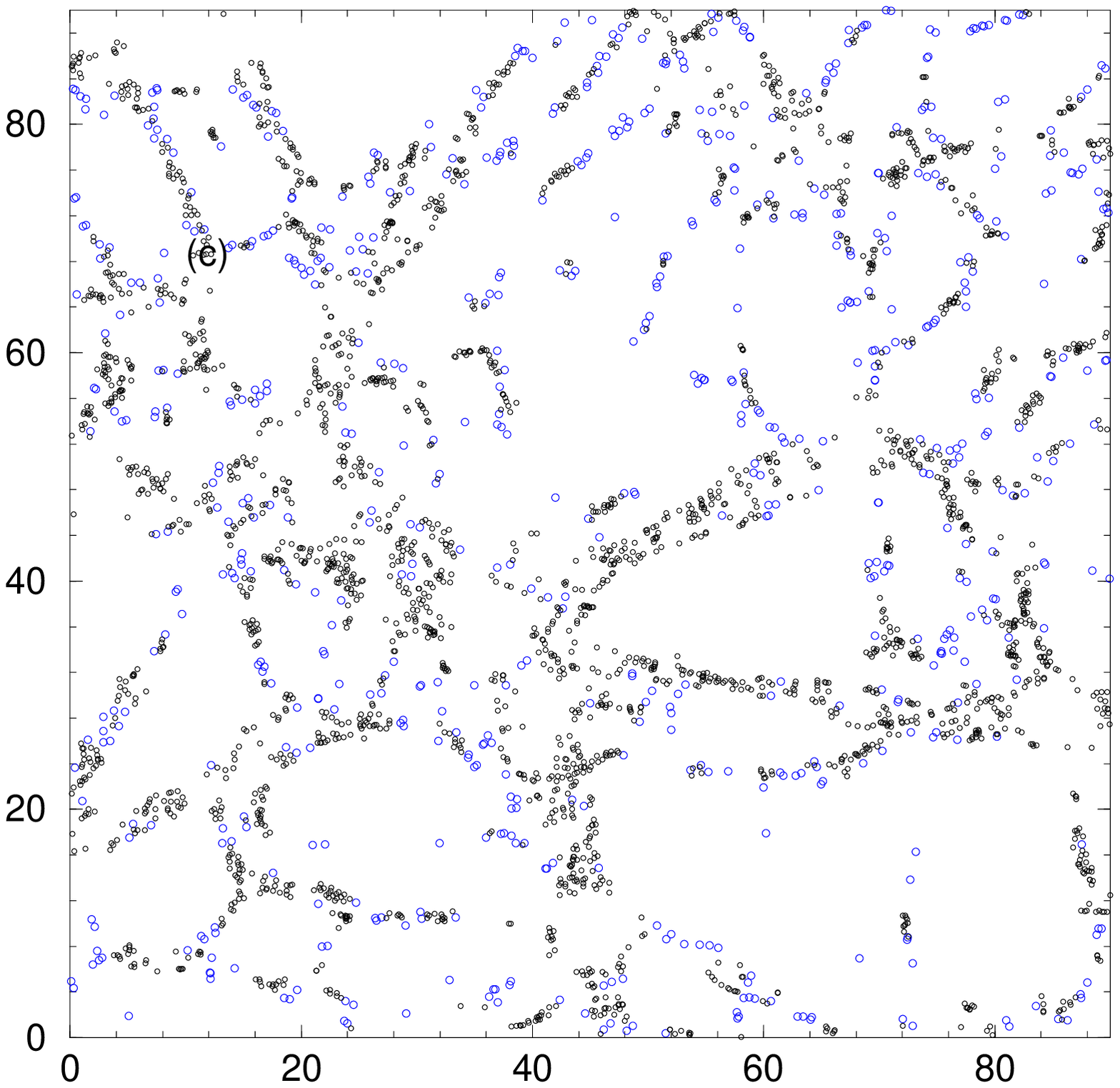}
\includegraphics{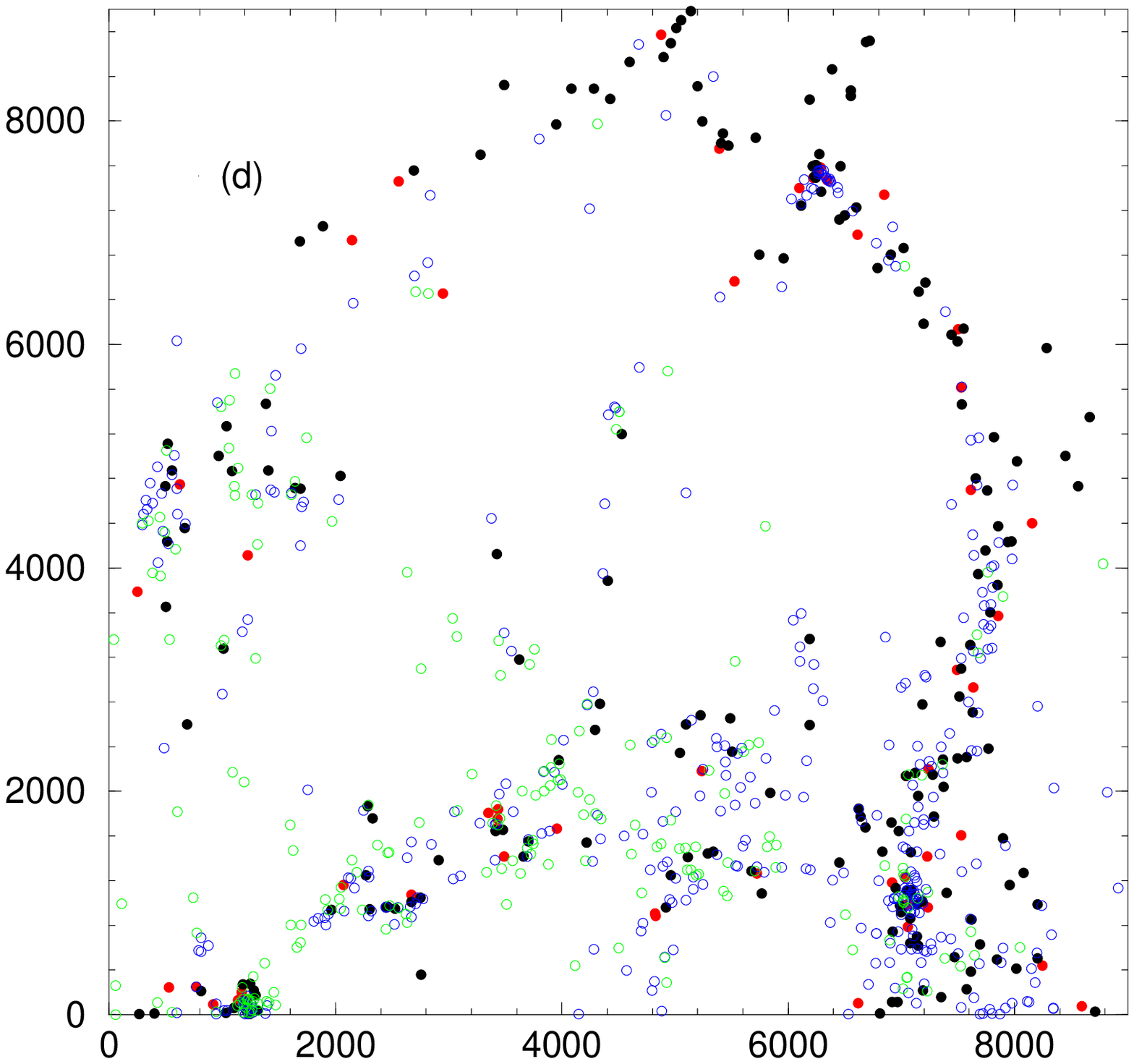}
\label{figure1}
\end{figure}

\subsection{The distribution of real and simulated galaxies}

In order to apply results of numerical simulations to samples of real
galaxies we must find the relationship between the distribution of
real and simulated galaxies (i.e. particles in the clustered
population). 

Based on considerations by EJS80 on the different evolution of under-
and over-dense regions, Einasto \& Saar (1986) divided particles in
simulations into void and clustered populations using the mean density
as the threshold density. In this case the topologies of simulated and
real galaxy samples are in very good agreement (Einasto \etal 1986).
A similar agreement between simulated and real galaxy samples exists
if one uses the correlation function test (Einasto, Klypin \& Saar
1986), the percolation and the filling factor tests for various
density levels (Einasto \etal 1986, Einasto \& Saar 1986, Gramann
1988, 1990).  This test was extended to the void diameter statistics
by Einasto, Einasto \& Gramann (1989), to the void probability
function by Einasto \etal (1991); and by Gramann \& Einasto (1991) and
GE92 to the power spectrum analysis.  A further step to check the
distribution of simulations with the real Universe was done by Gramann
\& Einasto (1991), Einasto \etal (1991), Frisch \etal (1995), and
Lindner \etal (1995) where not only simulated galaxies but also
simulated clusters were compared to real clusters.  In all these
studies a small smoothing length (about 1~\Mpc) was used to determine
the density field and to divide particles in simulations into the
high- and low-density populations.  These tests have shown that
statistical properties of simulated galaxies and clusters are very
close to properties of real galaxies and clusters. In other words, the
division of matter into the low-density primordial population in voids
and the clustered population with galaxies and clusters in
high-density regions describes well the actual distribution of
galaxies and clusters.

The next step in the comparison of simulations with the real Universe
was to investigate the possibility of using different threshold
densities to approximate the distribution of galaxies of different
morphology and luminosity.  It is well known that bright galaxies are
concentrated to central dense regions of groups, and that faint
companion galaxies are located in outskirts of groups (Einasto \etal
1974b).  Dressler (1980) extended the density relationship to
morphological types -- elliptical galaxies are located mostly in dense
regions and spirals in less-dense environments.  Einasto \etal (1991)
compared the void probability function for galaxy samples of different
luminosity limit with simulated samples selected at various threshold
density levels, and a similar comparison was made by GE92 using the
power spectrum test.  These studies have shown that void probability
functions, correlation functions and power spectra of simulated
galaxies selected using various threshold density intervals
approximate well the behavior of real galaxies of different luminosity
and morphology.

Now we shall compare the distribution of simulations with galaxies
using the 2-D simulation described above.  The analysis of the density
field in the Local Supercluster by E94 shows that the threshold
density, $\varrho_0$, which divides the non-clustered matter located
in voids and clustered matter associated with galaxies, is
approximately equal to the mean density.  Hydrodynamical simulations
of Cen \& Ostriker (1992) also indicate that the galaxy population is
located in regions of matter density above the average when smoothed
on 1 \Mpc\ scale. A more detailed hydrodynamical simulation by
Weinberg \etal (1997) shows that the distribution of gas particles of
different temperature is very different in regions of different local
density: the heating and successive cooling of gas occurs only in
over-dense regions.  We take the mean density as the threshold
density, $\varrho_0=1$.

In Figure~1 we present the distribution of void, field and cluster
particles in a (90~\Mpc)$^2$ region of the above simulation. We used a
threshold density $\varrho_0=1$ to separate void particles from
clustered ones, and $\varrho_{cl} = 5$ to separate field particles
from particles in clusters.  Panels (a), (b) and (c) show the
distribution of void, cluster, and field populations, respectively.
For comparison we show in panel (d) the distribution of real galaxies
in supergalactic coordinates in a sheet which crosses the Local, the
Coma and the southern corner of the Hercules supercluster.  We see
that simulated particles in voids are distributed rather uniformly,
while particles in the field are distributed along well-defined
filaments, and cluster particles form essentially spherical systems.

To check the possibility that $\varrho_0$ is different from 1 we have
compared the distribution of particles in the density intervals
$\varrho < 1$, $1 \leq \varrho < 1.5$ and $1.5 \leq \varrho <5$; see
Figure~1.  Particles with $1 \leq \varrho < 1.5$ form filaments in
less-dense environments and are absent in voids.  Their distribution
resembles the distribution of dwarf galaxies which form weak filaments
in super-voids, i.e. in voids defined by clusters of galaxies (see
panel (d), more detailed distributions are given in Figure~5 of
Lindner \etal 1995, and in Figure~3 of Lindner \etal 1996).  This
comparison shows that the use of density threshold to select various
simulated galaxies is well suited to discriminate cluster and field
galaxies, and among field galaxies to locate weak and massive
filaments.

This example shows that, at least for this simulation, the threshold
density values used reproduce well the actual distribution of galaxies
of different type.  Our Figure shows also that there exists no
one-to-one relationship between the distribution of simulated
particles selected in small threshold density intervals and real
galaxies chosen in small luminosity intervals.  The reason for the
absence of a very close relationship is clear: dwarf galaxies are
located also in clusters and in other high-density regions.  However,
mean statistical properties sensitive to the distribution of galaxies
in low-density environment (such as the void diameter distribution)
are rather similar for galaxy samples of various limiting absolute
magnitudes and for simulated galaxy samples using variable threshold
density levels (Lindner \etal 1995, 1996, Frisch \etal 1995).

\subsection{The distribution of matter and galaxies in different
environments} 

The previous analysis has shown a good agreement between the
distribution of simulated and real galaxies.  This analysis gives,
however, no answer to the question, how accurately galaxies follow the
distribution of matter in high-density regions: groups, clusters and
superclusters.  Direct observational data on the distribution of
galaxies and matter are needed to clarify this problem.

The distributions of the density of galaxies and matter in groups were
found to be essentially similar (Einasto \etal 1976, Vennik 1986,
Zaritsky \etal 1993, David \etal 1994, Pisani \etal 1995). Here we
ignore the difference in concentration of bright and faint galaxies in
groups, also we ignore the fact that within dark halos of galaxies the
distribution of baryonic and dark matter is different.  Since large
differences occur only on sub-Mpc scales, these differences do not
influence the power spectrum on scales of interest for the present
paper.

A comparison of the distribution of matter and light in clusters of
galaxies is possible using several of the indicators of the matter
distribution, like e.g.  gravitational lensing, X-ray emitting gas
distribution, or galaxy dynamics. These studies show that the matter
and luminosity distributions are rather similar, the concentration of
light is more pronounced than that of matter, similar to the
concentration of bright galaxies in groups (David \etal 1990, Carlberg
1994, B\"ohringer 1995, Squires \etal 1996, Carlberg \etal 1997,
Markevitch \& Vikhlinin 1997).  Such small differences can influence
the overall amplitude of the power spectrum as shown in the analysis
of results of numerical simulations with different threshold densities
and a sample of particles with positions shifted in high-density
regions (for details see the next subsection); the shape changes only
on scales comparable to the size of clusters (Figure~2).

The comparison of the distribution of matter and light in
superclusters is possible using numerical simulations.  Simulations
show that in large high-density regions (superclusters) more
primordial matter contracts to form clusters of galaxies than in
regions of lower density where systems of galaxies have lower
richness. This effect raises the amplitude of the power spectrum of
clusters, while the shape of the power spectrum changes only on
smaller scales (see Figure~2). To imitate this effect we have formed a
sample (sample ``Gal-120'' discussed below) where only a fraction of
particles in high-density regions is included.  An observational
argument in favor of the shape conservation of the power spectrum on
large scales is given by the similarity of power spectra of clusters
and galaxies in deep samples (see Figure~3 of Paper I for a comparison
of power spectra of clusters and that of the 3-D APM galaxy sample).

\subsection{Simulation of various galaxy populations}

\begin{table*}
\begin{center}
\caption[dummy]{Biasing parameters}
\begin{tabular}{lcccccll}
\hline
\hline
Sample &$\varrho_0$ & $N$ &$F_c$ & $b_F$ & $b_{mean}$ & $\delta_{bk}$ &
$\delta_{ba}$ \\
\hline

Matter	& 0	&262144	&1.0000	& 1.000& 1.000& 0.0\%&0.0\%	\\
Gal-1	& 1	&219965	&0.8391	& 1.192& 1.212& 0.4  &1.7 	\\
Gal-120	& 1	&215457 &0.8219 & 1.217& 1.167& 1.1  &4.1	\\
Gal-12	& 1 - 2	&202457	&0.7723	& 1.295& 1.374& 0.8  &6.1 	\\
Gal-2	& 2	&185206	&0.7065	& 1.415& 1.432& 1.1  &1.2	\\
Gal-2s	& 2	&185206 &0.7065 & 1.415& 1.428& 0.9  &0.9	\\
Clust	& 5	&135730	&0.5178 & 1.932& 1.861& 4.9  &3.7 	\\

\hline
\label{tab:bias}
\end{tabular}
\end{center}
\end{table*}

\begin{figure}[ht]
\vspace*{6.5cm}
\figcaption{ Left: Power spectra of simulated galaxies. The solid bold
line shows the spectrum derived for all test particles (the matter
power spectrum); dashed and dotted bold lines give the power spectrum
of all clustered particles (sample  Gal-1), and clustered galaxies in
high-density regions (sample Clust). Thin solid and dashed lines show
the power spectra of samples of particles with various threshold
densities and sampling rules (see Table~1 and text for details). Right:
the biasing parameter as a function of wavenumber, calculated from
definition eqn. (1). Samples and designations as in the left panel.
}
\includegraphics{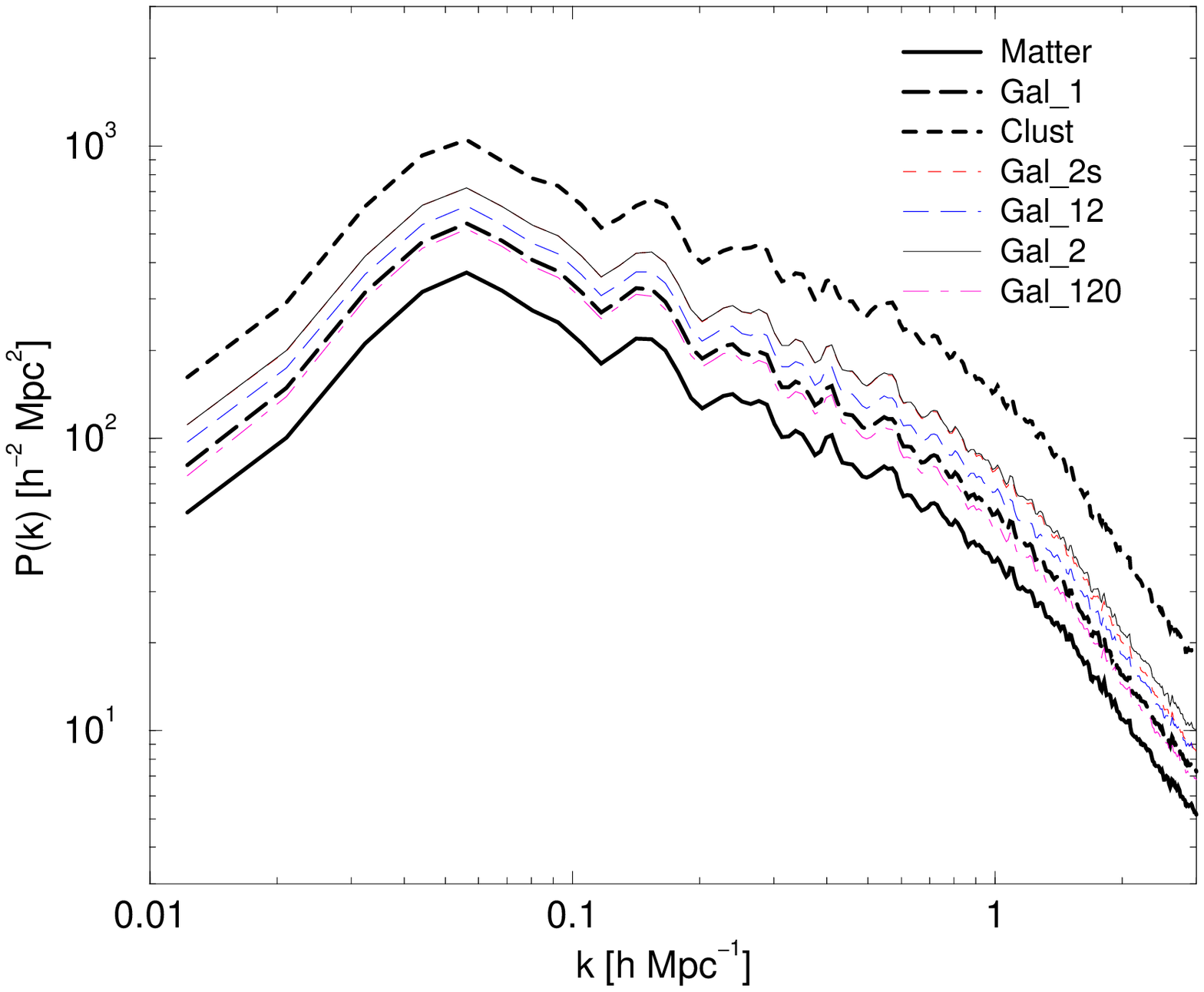} 
\includegraphics{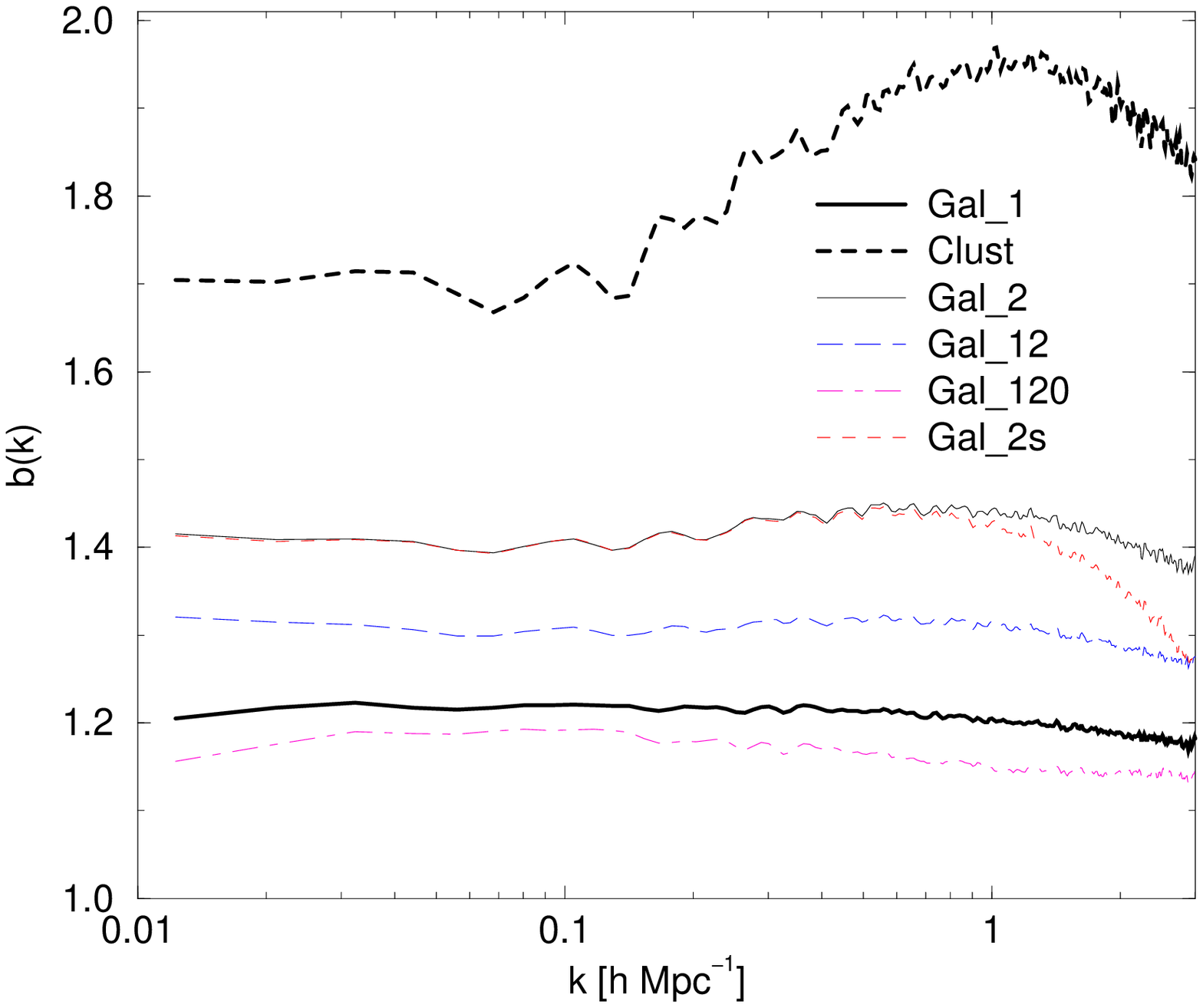}
\label{figure2}
\end{figure} 

For further tests we have formed a number of samples of particles with
various threshold density $\varrho_0$ and sampling rules. Table~1
gives the main parameters of these samples. $N$ is the number of
particles in samples, $F_c= N/N_{tot}$ is the fraction of particles in
the clustered population (a particular sample in units of the number
of particles in the sample of all matter); $b_F=1/F_c$ is the biasing
parameter calculated from eqn. (6); $b_{mean}$ is the mean value of
the biasing parameter found from the difference in the spectra of
matter and the sample (a mean value of local differences in the
wavenumber interval $0.01 < k \le 1.0$); $\delta_{bk}$ is the relative
error of the biasing parameter as a function of wavenumber (rms
deviation of the local biasing parameter value from the mean value
$b_{mean}$, in per cent); $\delta_{ba}$ is the relative error of the
biasing parameter $b_{mean}$ with respect to the theoretical value
$b_F=1/F_c$, in per cent.

Samples Gal-1, Gal-2 and Clust are defined by threshold densities,
$\varrho_0$, indicated in Table~1, such that all particles above the
threshold density are included. In sample Gal-12 a fuzzy threshold
density is placed randomly between densities 1 and 2, in sample
Gal-120 the threshold density is 1, but only 90\% of particles in very
high-density region ($\varrho \ge 20$) are included, i.e.\ this sample
imitates the deficiency of galaxies in superclusters. According to our
simulations particles in rich clusters of galaxies form 3 -- 5~\% of
the total number of particles; if the relative number of galaxies in
rich clusters is lower than in the field by a factor of up to 2 (in
clusters at least half of the baryonic matter is in the form of hot
X-ray gas, as indicated by direct observations and by the high value
of the mass-to-luminosity ratio in clusters), then this corresponds to
a $\sim 10$~\% decrease of the number of galaxies in the whole
population of high-density regions.  Finally, the sample Gal-2s
contains all particles above a threshold density of $\varrho_0=2$,
except that we added random shifts in an interval [$-0.25,0.25$] \Mpc\
to the positions of particles in high-density regions ($\varrho \ge
5$). This sample imitates possible difference in the concentration of
dark matter and galaxies in clusters (samples Gal-2 and Gal-2s
correspond to visible and dark matter in the clustered population,
respectively).

Figure~2 shows power spectra found for these samples. Obviously, all
power spectra of samples of clustered particles are similar to the
power spectrum of the matter, but have a higher amplitude. From the
difference in amplitude of the power spectra of these populations with
respect to the power spectrum of matter we derived the biasing parameter
as a function of wavenumber. The results are plotted in the right panel of
Figure~2. We see that for most samples the biasing parameter is almost
constant. Only for the cluster sample and the sample with particle
positions shifted in high-density regions (Gal-2s) the biasing
parameter on small scales deviates from the value observed on large
scales.

Table~1 shows that the biasing parameter $b_{mean}$, found from the
difference in power spectra, is surprisingly close to the value
expected from the number of particles in respective samples, $b_F$.
These calculations show that eqn. (5), derived for the case of
constant density of the void population, works well even in cases when
we consider cluster galaxies. In this case the population of particles
in low-density regions contains not only real void particles, but also
particles which imitate field galaxies in filaments. In other words,
eqn. (5) is very robust and insensitive to the distribution of the
low-density population, and its relative error is a few per cent. Only
in case of the sample Gal-120 where part of the galaxies in
high-density regions have been removed, the biasing parameter is about
5~\% lower than expected from eqn. (5).  Similarly, in the case of the
sample with a fuzzy threshold the mean biasing parameter value is
about 5~\% higher than predicted from eqn. (5).

These data indicate that there is no evidence for the presence of
large differences in the shape of the power spectra of matter and
galaxies on scales of interest for the present study. Differences
directly observed between cluster and galaxy samples, and predicted
from small variations in the concentration of galaxies and matter in
clusters and superclusters, are confined to smaller scales only, $k \ge
0.5$~\hmpc.

\vskip0.5cm

The principal result of our analysis is that power spectra of the
population of all clustered particles and matter have similar shape.
The difference in amplitude of power spectra of clustered particles
and matter is given by the fraction of matter in the clustered
population.  Analysis done for various 3-D models of structure
formation since the study of GE92 has reached the same conclusion
(Frisch \etal 1995).  Sampling peculiarities which imitate various
galaxy populations change the mean biasing parameter only very
modestly.

\section{The amplitude of density fluctuations}

The previous analysis has shown that the power spectrum of the
clustered population (galaxies) can be reduced to the power spectrum
of matter using a simple formula (5), if we know the fraction of matter
in voids and in high-density regions.  In principle, the amount of matter
in voids can be calculated using the velocity field and applying
methods of the restoration of the matter distribution (for a recent
analysis see Freudling \etal 1998).  The accuracy of peculiar velocity
measurements is, however, not sufficient to get reliable results.  For
this reason we use a different approach here and we derive the fraction of
matter in voids and in the clustered population from numerical
simulations.  In doing so we assume that on large scales the evolution
of the structure is determined by gravity alone, and that simulated
galaxy populations with appropriate threshold densities approximate
real galaxy populations.

From numerical simulations it is straightforward to find the
distribution of particles as a function of the local density of their
environment. A simple counting of particles with associated density
values exceeding the threshold $\varrho_0$ yields the fraction of
matter in the clustered population, $F_c$. The problem is how to
identify the present epoch in the simulation.

During the evolution of the Universe matter flows from low-density
towards high-density regions, and the fraction of matter in the
clustered population, $F_c$, grows. Simultaneously the amplitude of
the power spectrum increases; the mean amplitude can be expressed in
terms of density fluctuations in a sphere of radius $r=8$~\Mpc,
$\sigma_8$.  We see that there exist a relation between $F_c$ and
$\sigma_8$; thus the epoch of the simulation can be measured in terms
of the $\sigma_8$ parameter.  If the present value of $\sigma_8$ of
matter is known from other sources then the whole simulation can be
calibrated.

E94 determined rms density fluctuations in the Local supercluster on
galactic scales, $\sigma_{1.2}$, and used this calibration to fix the
present epoch of simulation. They found that for a wide class of models
(with and without cosmological constant) the present fraction of matter in
voids is almost independent of the model.  The fraction of clustered
matter is $F_c=0.85 \pm 0.05$. 

Here we modify the method of E94. The problem lies in the following:
from simulations we know the amplitude of fluctuations of matter
whereas from observations we have the amplitude of fluctuations of
galaxies. These two quantities are related through a formula similar to
(5).  $\sigma^2(r)$ is proportional to $P(k)$, thus we get the
relation between $(\sigma_8)_{gal}$ and $(\sigma_8)_{m}$
$$
(\sigma_8)_m = F_{gal} (\sigma_8)_{gal}.
\eqno(7)
$$
Here we assume that $F_{gal}=F_c$.  This formula holds under the same
assumptions as eqn. (5). It is practically exact for the whole galaxy
population (see the error analysis given in Section 2.8).

Eqn. (7) gives one relation between $F_c$ and $(\sigma_8)_{m}$,
another relation can be found from numerical simulations (see
below). By simultaneous solution of both relations we can find both
parameters for the present epoch, $F_c$ and $(\sigma_8)_{m}$. Eqn. (6)
yields then the bias factor of the clustered matter (galaxies).

\subsection{The amplitude of galaxy density fluctuations}

The rms amplitude of density fluctuations of galaxies,
$(\sigma^2(r))_{gal}$ in a sphere of radius $r$, is a direct
observable.  Usually it is determined from counts in cells or through
the correlation function using a power-law approximation of the
correlation function (Davis \& Peebles 1983). For a recent
determination of $(\sigma_8)_{gal}$ see Willmer, da Costa \&
Pellegrini (1998).

Here we apply a different method to determine $(\sigma^2(r))_{gal}$.
The rms amplitude of density fluctuations in a sphere of radius $r$ may
be found by integrating the power spectrum:
$$
\sigma^2(r)= {1\over2\pi^2}\int_0^{\infty} P(k)
W^2(kr) k^2 dk,
\eqno(8)
$$
where $W(kr)$ is the window function. We shall use a top-hat window
$$
W(kr) = {3(\sin kr - kr \cos kr) \over 
(kr)^3}.
\eqno(9)
$$

The function $\sigma^2(r)$ is an integral representation of the power
spectrum.  Like all integral functions it is less dependent on local
irregularities than the integrand. We shall determine this function 
from the observed power spectrum of all galaxies.

Observations allow to determine the power spectrum in the wavenumber
interval $0.03 \leq k \leq 1$ (Paper I).  To apply the eqn. (8) we
have to extrapolate the observed spectrum to larger and smaller
scales.  For this extrapolation we shall use theoretical model spectra
which fit the observed spectra. On small scales the observed
non-linear power spectrum was reduced to a linear one (see Einasto
\etal 1999b, Paper III for details).  The observed power spectrum was
determined for two populations, representing galaxy samples of the
Universe which include high-density regions and medium-density
regions, $P_{HD}(k)$, and $P_{MD}(k)$, respectively.

In Figure~3 we show the function $\sigma(r)$ calculated for both
variants of the power spectrum, $P_{HD}(k)$ and $P_{MD}(k)$.
Theoretical spectra were calculated for a Hubble parameter $h=0.6$,
baryonic density parameter $\Omega_b=0.04$, density parameter
$\Omega_0=0.4$ and cosmological constant parameter
$\Omega_{\Lambda}=0.6$. In order to test how the value of $\sigma_r$
is affected by the extrapolation, we varied the density parameter and
associated cosmological constant for a flat model (see figure caption
for details).  We see that all variants of $\sigma(r)$ coincide around
the scale $r=8$~\Mpc. This proves that $\sigma_8$ is almost
insensitive to the details of the extrapolation of the power spectrum
on small and large scales, and to the exact shape of the power
spectrum around the maximum.  A similar conclusion was obtained by
White, Efstathiou \& Frenk (1993). For the galaxy power spectrum
$P_{HD}(k)$ we obtain
$$
(\sigma_8)_{gal} = 0.89 \pm 0.05,
\eqno(10)
$$
which is rather close to the often used value of unity.  The error of
$(\sigma_8)_{gal}$ is determined by the error of the amplitude of the
observed mean galaxy power spectrum, due to the scatter of power
spectra of various samples. If we add the possible systematic error
due to the uncertainty in the overall normalization of the amplitude,
we get for the $1~\sigma$ error of $(\sigma_8)_{gal}$ 10~\%, i.e. $\pm
0.09$.

\begin{figure}[ht]
\vspace*{7cm}
\figcaption{Integrated power spectrum, $\sigma(r)$. Bold solid and
dashed lines give $\sigma(r)$ calculated from power spectra
$P_{HD}(k)$ and $P_{MD}(k)$, respectively. Thin lines show linear
extrapolations of $\sigma(r)$ calculated using flat CDM models with
cosmological constant parameter $\Omega_{\Lambda} = 0.2, \dots 0.8$,
Hubble constant $h=0.6$ and baryonic density parameter
$\Omega_b=0.04$.  All functions are reduced to matter using a bias
factor $b=1.18$ (E94).  } 
\includegraphics{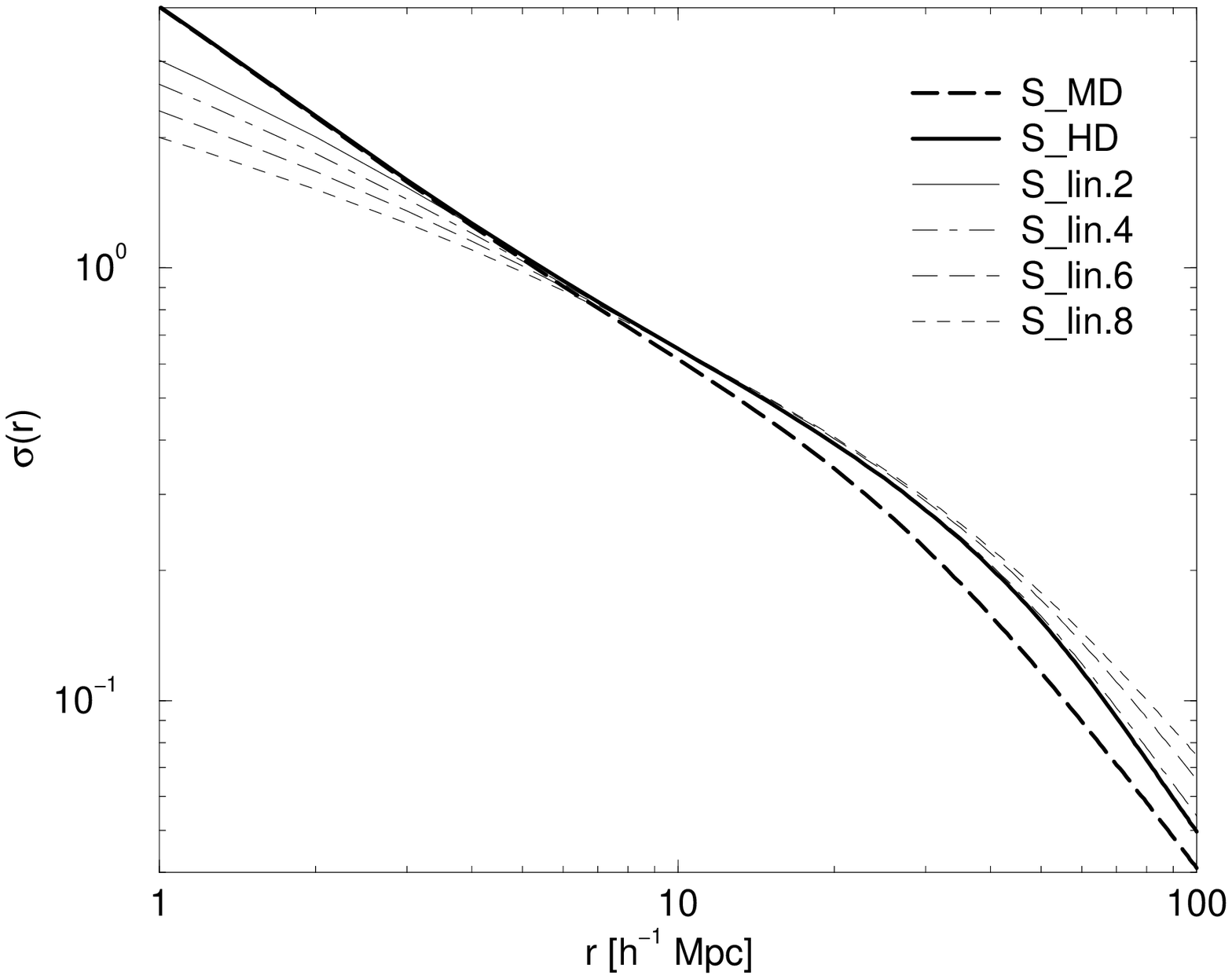}
\label{figure3}
\end{figure}

\begin{table*}
\begin{center}
\caption[dummy]{Simulation parameters}
\label{tab:param}
\begin{tabular}{lccccccc}
\hline
\hline
Model   & Number    & Number  & $L_{box} $ &$\Omega_{0}$&$\Omega_{\Lambda}$
& $h$ & $(\sigma_8)_{max}$ \\
        & of particles&of cells & (\Mpc) & \\
\hline
SCDM   & $128^{3}$ & $128^{3}$  & 200 & 1.0 & 0.0 & 0.5 & 1.16   \\
LCDM1  & $128^{3}$ & $128^{3}$  & 280 & 0.3 & 0.7 & 0.7 & 1.07   \\
LCDM2  & $128^{3}$ & $128^{3}$  & 280 & 0.3 & 0.7 & 0.7 & 1.00   \\
OCDM   & $128^{3}$ & $128^{3}$  & 280 & 0.5 & 0.0 & 0.7 & 0.91   \\

\hline
\label{tab:prop}
\end{tabular}
\end{center}
\end{table*}

\subsection{Reduction to matter power spectrum}

Eqn. (7) contains one observed quantity, $(\sigma_8)_{gal}$, and 2
unknowns, $(\sigma_8)_m$ and $F_{gal}$. To find a solution we need one
more relation between these two unknowns. Here we use the fact that
$\sigma_8$ grows with time and is suited to measure the epoch in
numerical simulations.  We recall that $F_c=F_{gal}$ is the total
fraction of matter in high-density regions where the local density
(smoothed on Mpc scale) exceeds the mean density.  Initially, by
definition of the mean density, we have $F_c= 0.5$. During the evolution
matter flows from low-density regions towards high-density regions
which evolve to clusters, groups and galaxy filaments, thus $F_c$
increases.  

To determine the relation between $F_c$ and $\sigma_8$, we made 3-D
simulations with a standard CDM model (SCDM), two realizations of a
spatially flat model with cosmological constant (LCDM), and an open
model (OCDM).  Simulations have been made with the P$^3$M code of
Couchman (1991); simulations were run until $\sigma_8 \approx
1$. Model parameters are given in Table~\ref{tab:prop}.  Simulations
have been made with two sets of initial positions of particles. In the
first case particles were placed on a regular grid and then displaced
via the Zeldovich approximation. In the second case initial particle
positions have a homogeneous glass-like distribution; these positions
are then used for displacing the particles according to the Zeldovich
approximation, using the same random phases as in the first case.
Homogeneous distributions as input for initial conditions are better
because they inhabit no structure like a grid.  Therefore no remnants
of the grid are seen as the simulation evolves.

As discussed above, we assume that in high-density regions ($\varrho
\geq \varrho_0=1$) the distribution of the total matter density is
identical with the distribution of the number density of galaxies (the
density $\varrho_0$ is expressed in mean density units).  We calculate
the density on the location of galaxies using an adaptive smoothing
algorithm: it is determined from the outer radius of a sphere which
contains 12 nearest neighbors to the galaxy. In systems of galaxies
this number corresponds approximately to a smoothing scale comparable
to the size of typical systems of galaxies -- clusters, groups and
filaments.

\begin{figure}[ht]
\vspace*{8cm}
\figcaption{Left: evolution of the integrated density distribution for
the OCDM model. The fraction of matter in low-density regions is
plotted for different epochs, indicated by the $\sigma_8$ parameter.
Right: relation between the fraction of matter in galaxies, $F_{gal}$,
and $\sigma_8$. Thick bold solid line shows the relation from
eqn. (5), bold solid, dashed and dot-dashed lines give the relation
obtained from numerical simulations of how voids are emptied in 
different cosmological models (see Table~\ref{tab:prop}).  Models with
glass-like initial conditions are plotted as bold lines, models with grid
initial conditions as thin lines. The mean error of $F_{gal}$ due to
uncertainty in the threshold density level is $\approx 0.05$.
}
\includegraphics{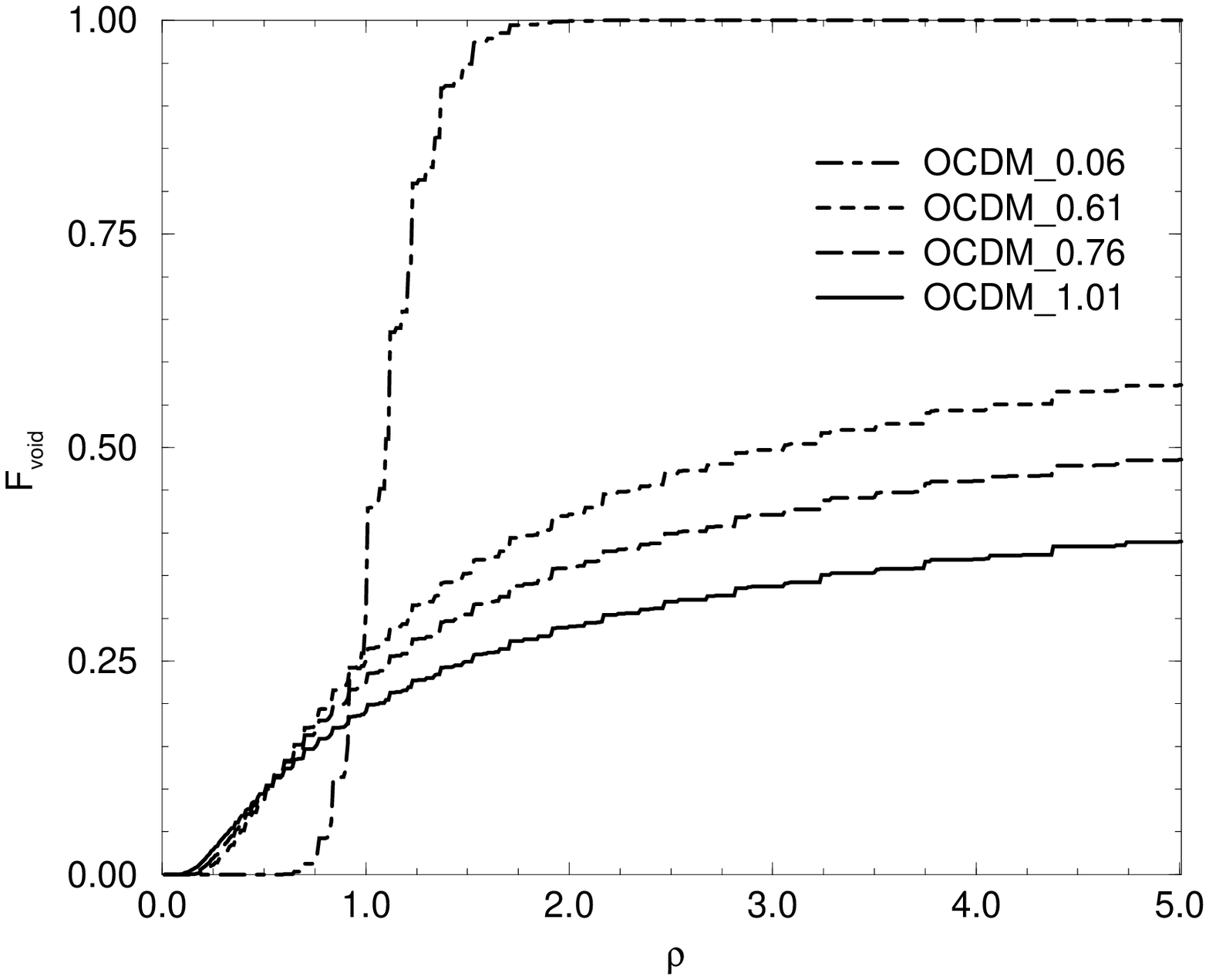}
\includegraphics{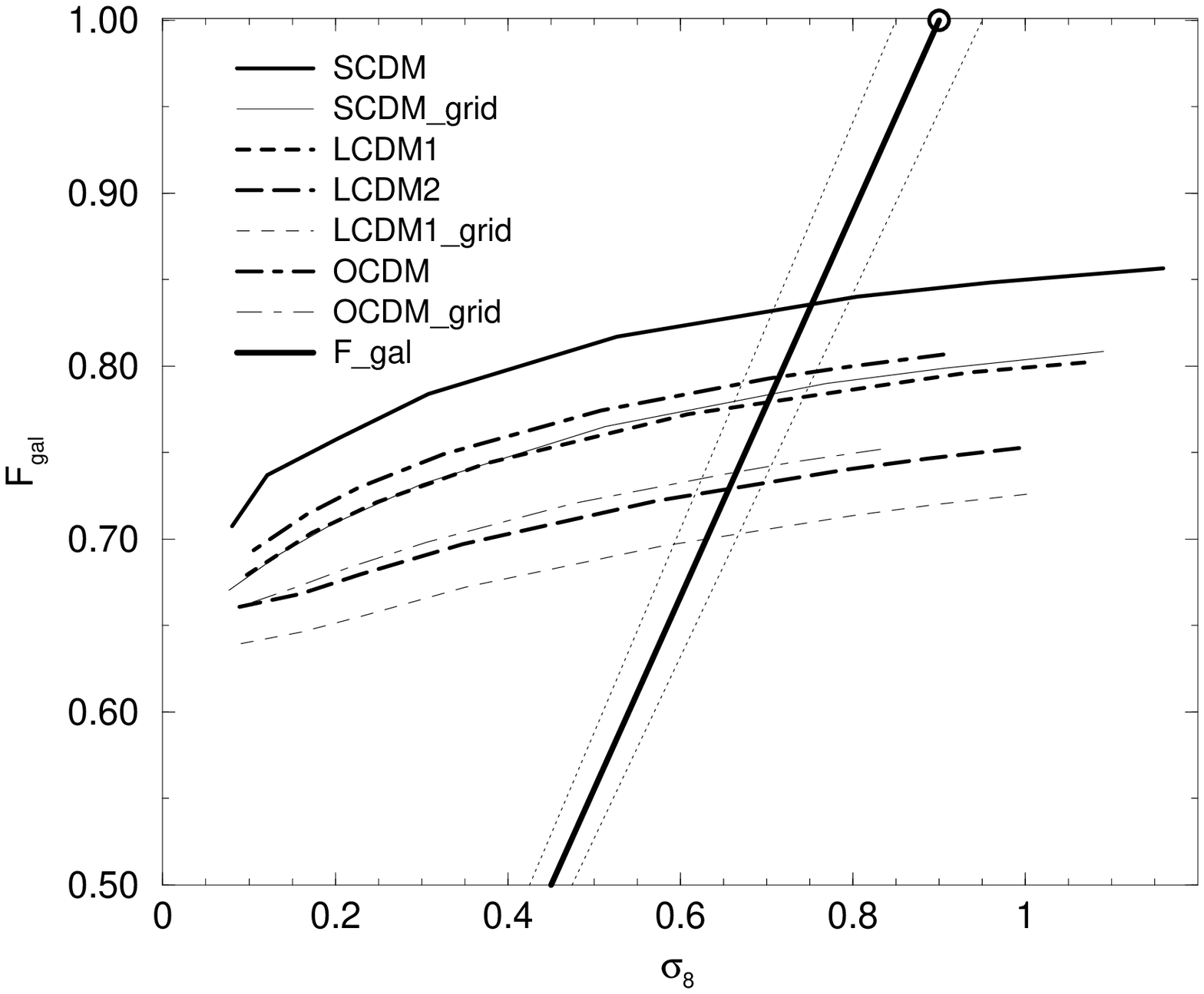}
\label{figure4}
\end{figure}

For all time-steps we calculated the integrated density distribution,
i.e. the fraction of particles located in regions with density $F(\leq
\varrho)$ for different $\varrho$.  For the OCDM model, our results
are shown in Figure~4a. They are very similar to those plotted in
Figure~3 of E94. The fraction of matter in galaxies is given in
Figure~4b, expressed in terms of $\sigma_8$.

The relation between $F_c$ and $\sigma_8$, following from the eqn.
(7), is also shown. For a given model of structure evolution, these
two relations fix both parameters.  The void evacuation is model
dependent, thus we have a different solution for each model. In LCDM
and OCDM models voids are emptied and matter flows into dense regions
at surprisingly similar speeds while for SCDM void evacuation occurs
faster. Also we see that models with grid-like initial conditions
yield systematically lower values for $F_{gal}$. As glass-like initial
conditions give a smoother density field we prefer to use these
models.  For the SCDM model we obtain $F_{gal}=0.83$ and
$(\sigma_8)_m=0.75$, for the LCDM1 model $F_{gal}=0.78$ and
$(\sigma_8)_m=0.70$, for the LCDM2 model $F_{gal}=0.73$ and
$(\sigma_8)_m=0.66$, and for the OCDM model $F_{gal}=0.79$ and
$(\sigma_8)_m=0.72$.

We see that the amplitude fluctuation parameter, $(\sigma_8)_m$,
depends slightly on the density parameter of the Universe. Independent
determinations (Ostriker \& Steinhardt 1995, Bahcall, Fan \& Cen 1997,
Bahcall \& Fan 1998) favor a low-density Universe, thus we prefer to
use results obtained for LCDM and OCDM models which yield $F_{gal} =
0.75 \pm 0.08$ for the present epoch, in good mutual agreement. For
the biasing parameter of galaxies relative to the matter, $b_{gal}$,
we obtain:
$$ 
b_{gal}=1/F_{gal} = 1.32 \pm 0.13. 
\eqno(11)
$$ 
We see that the fraction of matter in galaxies according to new models
is smaller than found by E94 which leads to a larger value of the
biasing parameter. The difference is partly due to differences in
models (most models used by E94 were standard CDM), and partly due to
differences in the methods used in calculations and fixing the present
epoch.

\section{Discussion}

\subsection{The biasing of galaxies relative to matter}

We have investigated the biasing of galaxies relative to matter.  We
use the term ``biasing'' to denote the difference between the
distribution of the whole matter and the matter associated with
galaxies. Already in early stages of cosmological studies it was clear
that the evolution of density perturbations in under- and over-dense
regions is completely different (Z70), which explains the presence of
voids since matter in low-density regions cannot contract and form
galaxies (JE78, EJS80).  Thus the discovery of voids was a clear
indication for the dominating role of the gravity in the evolution of
the Universe on large scales.  This leads us to our basic assumptions
that the evolution of the structure on scales of interest is due to
gravity, and that density fluctuations are Gaussian and adiabatic.
These assumptions have two important consequences.  First, the
evolution of under- and over-dense regions is completely different:
the density in under-dense regions decreases (approximately
exponentially) but never reaches zero; the density in over-dense
regions increases until the matter collapses (Z70, EJS80).  Collapsed
regions form a web of intertwined filaments and knots (JE78, Einasto
\etal 1983, Melott \etal 1983, Bond \etal 1997, Cen \& Simcoe 1997).
Second, in order to form a galaxy or a system of galaxies, the clump
of primordial matter must have a density exceeding some critical limit
in a given volume (Press \& Schechter 1974). For these reasons the
galaxy formation is a density threshold phenomenon.

Dynamical studies of galaxies and clusters have shown that the
dominating population in the Universe is dark matter which forms halos
around galaxies, groups and clusters of galaxies.  Practically all
massive galaxies are located in groups or clusters; the brightest form
main galaxies of groups, other group members are dwarf companion
galaxies (Einasto \etal 1974b, Zaritsky \etal 1993).  The
characteristic size of groups of galaxies is of the order 1~\Mpc\
(Einasto \etal 1984), thus, in order to find the {\em true} density
field of matter in the Universe the discrete distribution of galaxies
(and particles in numerical simulations) is to be smoothed using a
smoothing scale of the order of the size of groups, i.e. about 1~\Mpc\
(Einasto \& Saar 1986).

Conventionally the biasing is defined through the local density
contrast, which is calculated from a smoothed density field using a
smoothing scale of the order of 10~\Mpc.  Excessive smoothing mixes
unclustered DM in voids and clustered DM in high-density regions which
makes the simple biasing phenomenon rather complicated.  To avoid
excessive smoothing we define the biasing parameter using the
difference in power spectra of populations, in our case of the
population of all galaxies with respect to matter.  We apply no
additional smoothing, i.e.\ power spectra are calculated from particle
positions, and the density field is calculated by interpolation of
these positions on a grid which has a scale of the same order as real
systems of galaxies.  We use density also to find the population
membership of particles, either void particle in low-density regions
or simulated galaxy (with dark halo) in high-density regions.  Here
again we interpolate particle positions within the grid.

In the determination of the density at the location of particles we
tacitly assume that DM is non-baryonic and consists of particles of
small mass. Under this assumption we can consider DM as a fluid which
has a continuous density field. As shown by dynamical observations and
numerical simulations, DM forms density enhancements around galaxies
and in clusters of galaxies. The distribution of luminous matter
differs from the distribution of DM on galactic and cluster
scales. There exists no indication for large-scale segregation between
luminous and dark matter. We have investigated the influence of
possible small differences in the spatial distribution of luminous and
dark matter to power spectra of galaxies and matter. Our results show
that on scales of interest for the present study these differences are
small or negligible.

Our study demonstrates that the main difference between power spectra of
galaxies and matter is the amplitude only, which is determined by
the fraction of particles in high-density regions. Small differences
in the spatial distribution and in the density threshold are the
reason for cosmic scatter of the biasing parameter around the value
defined by the fraction of particles in high-density regions; this
error in the density threshold is of the order of 10~\% or less. The
density distribution (Figure~4a) shows that a 10~\% error in the
threshold density leads to a 3~\% error of the fraction of matter in
the clustered population.  Biasing parameter values $b \le 1$ are
possible only for samples with a very large deficit of particles in
high-density regions (see Figure~2 of Paper I). Such a large deficit is
not supported by observation, as the mass-to-luminosity ratio of
groups of galaxies is approximately the same as in clusters. A small
deficit of luminous matter in rich clusters (high mass-to-luminosity
value) is simulated in our test sample Gal-120: its power spectrum has
a lower amplitude as defined by the number of particles, but only by a
few per cent. This adds a component to the cosmic scatter of the
biasing parameter.

The main lesson from this study is that the gravitational origin of the
structure evolution poses strict limits to the biasing parameter. It
is determined by the fraction of total matter in low- and
high-density regions. As matter flows continuously away from
low-density regions, the fraction of matter in high-density regions
remains between 0.5 (the initial value) and 1 (the limit in very far
future) during the whole evolution of the structure. Thus the   
corresponding biasing parameter of galaxies relative to matter lies
between 2 (initial value) and 1 (limit in the future). These values
apply if galaxies exactly follow the distribution of particles in
high-density regions. Our analysis has shown, that differences of the
distribution of galaxies and matter in high-density regions may change
these theoretical biasing parameter values by up to ten per cent.

The second lesson learnt is that both the amplitude (i.e. the biasing  
parameter) {\it and} the shape of the power spectrum change very little  
due to possible disturbing effects. The amplitude is changed
considerably only in the case when galaxy samples are incomplete in
high-density regions.  The shape is changed by differences in the
concentration of matter and galaxies in groups, clusters and
superclusters, but this affects the power spectrum on small scales
only (see Table~1 and Figure~2) which is of less importance for the
present study.

We have followed this approach to the biasing problem motivated by the
discovery of the difference in the distribution of galaxies and matter
(JE78, EJS80). It is a bit surprising that in the majority of studies on
this subject the problem has been complicated by smoothing over large
scales. The latter distorts the distribution of void particles and the
clustered matter.

\subsection{The amplitude of density fluctuations}

We have studied the evacuation of voids and the concentration of
matter to high-density regions through numerical simulations.  We used
the rms density fluctuation on a 8~\Mpc\ sphere, $\sigma_8$, as a
parameter which characterizes the epoch of the simulation. We
determined the present value of this parameter from the power spectrum
of galaxies, $(\sigma_8)_{gal}=0.89 \pm 0.09$. We find two relations
for $(\sigma_8)_m$ and $F_c$, and derive values for both parameters.
We find the fraction of matter in the clustered population $F_c = 0.75
\pm 0.08$, and $(\sigma_8)_m=0.68 \pm 0.09$. These data yield a value
$b_{gal}=1.32 \pm 0.13$ for the bias parameter of all galaxies
relative to matter.

The amplitude of matter density fluctuations can be fixed on two
different scales.  COBE data measure the amplitude of density
fluctuations on very large scales (e.g. Bunn \& White 1997).  On
galactic scales the amplitude of density fluctuations is quantified by
the $\sigma_8$ parameter.  The problem here is how to link the
distribution of galaxies to that of matter. To avoid this difficulty
White, Efstathiou \& Frenk (1993) used an indirect method, based on
the cluster abundance (number density).  Cluster abundance depends on
the amplitude of density fluctuations and density parameter, $\sigma_8
\Omega_0^{0.6}$.  Previously, the $\sigma_8 - \Omega_0$ relation was
found as $\sigma_8\sim 0.50 \Omega^{-0.41\pm 0.02}$ (Eke, Cole \&
Frenk 1996, Viana \& Liddle 1996, Pen 1996, Cen 1998).  Bahcall, Fan
\& Cen (1997), Fan, Bahcall \& Cen (1997) and Carlberg \etal (1997)
point out the fact that the evolution of cluster abundance depends
strongly on $\sigma_8$, making it possible to determine both
parameters separately.  Combining the observed abundance of local rich
clusters (Bahcall \& Cen 1992, 1993) with cluster evolution data
yields the following parameters: $\Omega_0=0.3 \pm 0.1$,
$\sigma_8=0.83 \pm 0.15$, and $b_{gal}=1.2 \pm 0.2$. Using a similar
method Eke \etal (1998) find $\Omega_0=0.44 \pm 0.2$, $\sigma_8=0.67
\pm 0.1$ for an open Universe, and $\Omega_0=0.38 \pm 0.2$,
$\sigma_8=0.74 \pm 0.1$ for a flat Universe with a cosmological
constant. Bahcall \& Fan (1998) find $\Omega_0=0.2^{+0.3}_{-0.1}$
based on similar arguments.  A simple estimate of the cluster mass in
the comoving volume from which that mass originated yields a
mass-density estimate of the Universe $\Omega_0=0.24 \pm 0.10$
(Carlberg \etal 1996).  These results suggest that the standard CDM
model can be excluded at a confidence level of more than 99\%.

For comparison we note that COBE normalization provides
$(\sigma_8)_m=1.2 \pm 0.1, \quad 1.0 \pm 0.1 \quad \mbox{and} \quad
0.9 \pm 0.1$ for SCDM, LCDM, and OCDM, respectively.  As
$(\sigma_8)_m$ calculated from COBE normalization is an extrapolation
from 1000~Mpc to 10~Mpc scales, based on certain models of
structure evolution, the (dis)agreement between the direct
determination of $\sigma_8$ and COBE extrapolation to this scale can
be considered as evidence for the quality of the model used.

Within the errors our value of the $\sigma_8$ parameter coincides with
those suggested by Carlberg \etal (1997), Bahcall, Fan \& Cen (1997)
and Eke \etal (1998). A recent determination of the density parameter
and $\sigma_8$ by Bahcall \& Fan (1998) from the abundance of rich
clusters of galaxies at high redshifts yields a rather high value,
$\sigma_8=1.2^{+0.5}_{-0.4}$; however, the error is large.  Since we
used completely different input data, our determination is independent
of previous ones. Moreover, our method uses directly observed power
spectra and simple gravitational physics of void evacuation, thus the
danger of the presence of large systematic errors is small.  We may
conclude that this parameter is now known rather reliably.

Using the integrated power spectrum we can also determine the Excess
Power parameter, $EP=3.4 \sigma_{25}/\sigma_8$, introduced by Wright
\etal (1992). We find $EP=1.42 \pm 0.05$ and $EP=1.28 \pm 0.05$, from
the power spectra $P_{HD}(k)$ and $P_{MD}(k)$, respectively. Both
values are close to the value $EP=1.30 \pm 0.15$ found by Wright et al.
A similar power spectrum shape parameter was introduced by Borgani
\etal (1997).

The error of $(\sigma_8)_m$ depends on the errors of
$(\sigma_8)_{gal}$ and $F_{gal}$.  We have no reason to believe that
the amplitude of the observed power spectrum of galaxies has a
considerable systematic error. Thus we can accept the quoted error of
$(\sigma_8)_{gal}$ as a realistic one.  The fraction of matter in
galaxies derived from numeric simulations is less certain. But we can
estimate upper and lower limits for this quantity.  EJS80 and E94 have
found analytic approximations for several simple scenarios for the
evacuation of voids (planar and spherical void models).  Negative and
positive density fluctuations grow simultaneously, and the present
fraction of matter in voids (and in the clustered population) depend
on the effective epoch of structure formation (collapse),
$z_{form}$. We find (see Figure~7 of E94) that for the linear void and
wall model we obtain $0.6 < F_c < 0.8$, if $1 < z_{form} < 5$, while
spherical void models give $0.8 < F_c < 0.9$.  As a collapse at very
early and very late epochs can safely be excluded, these simple models
suggest that the present value of the fraction of matter in the
clustered population must lie in the interval $0.6 < F_c < 0.9$. Using
these limits we get $0.54 < (\sigma_8)_m < 0.8$. Our accepted value
$(\sigma_8)_m=0.68$ lies just in the middle of this interval. If we
consider the limits derived from these simple analytic models as
$3\sigma$ errors, we get for $1\sigma$ error 0.05, in good agreement
with the error estimate calculated from formal errors of parameters
used to find $(\sigma_8)_m$.

To conclude the discussion we stress again that $(\sigma_8)_m$
characterizes the rms amplitude of density fluctuations on galactic
scales, similar to COBE observations which fix the amplitude of
density fluctuations on a scale of $\sim 1000$~Mpc. Direct observables
are the rms galaxy density fluctuations at the present epoch,
$(\sigma_8)_{gal}$, and rms temperature fluctuations at the
recombination epoch, respectively.  In both cases, matter density
fluctuations at the present epoch are calculated using theoretical
models which involve simple physics, through the evacuation of voids
and the growth of the amplitude of density fluctuations, respectively
for $(\sigma_8)_m$ and COBE cases. These methods to calibrate density
fluctuations on different scales are complementary and independent,
i.e.\ a correct model must pass both checks.

\section{Conclusions}

In cosmological studies the linear bias factor is often defined
through the $\sigma_8$ parameter ($b \equiv 1/\sigma_8$, see Brodbeck
\etal 1998). Actually these quantities are independent parameters, the
bias parameter characterizes the difference in amplitude of power
spectra of galaxies and matter, and the $\sigma_8$ parameter the
present amplitude of density fluctuations of matter on galactic
scales.  The main goal of this paper was to elaborate methods to
determine these two parameters, and to apply  methods using
actual data.

Our approach to the biasing phenomenon is based on the observation
that there exist large voids in the galaxy distribution. This is a
clear indication that the evolution of the structure in the Universe
is primarily due to gravity.  Further we assume that primordial
density fluctuations are Gaussian and adiabatic.  We have shown that
under these assumptions the formation of galaxies is a threshold
phenomenon, i.e. that in under-dense regions galaxies do not form at
all, and that in over-dense regions galaxies and matter are
distributed very similarly (ignoring differences on galactic scales).

Our first conclusion from the biasing analysis is that all
matter in the Universe is divided into two main populations, the
unclustered primordial matter in voids and the clustered matter in
high-density regions associated with galaxies.  Our analysis and
high-resolution hydrodynamical simulations of galaxy formation show
that the threshold density which divides the unclustered matter in
voids and the clustered matter associated with galaxies, is
approximately equal to the mean density of matter, if smoothed on
scales comparable to the characteristic scale of groups of galaxies
(about 1~\Mpc).  Using higher threshold densities it is easy to select
galaxies located in filaments, while a still higher threshold density
corresponds to galaxies in groups and clusters.  Using intermediate
threshold density intervals it is even possible to simulate
statistically populations of galaxies of different luminosity.

We have investigated the influence of the density threshold to power
spectra of galaxies and clusters of galaxies.  The population of
clustered particles is derived from the population of all particles by
the exclusion of particles located in low-density environments.  Our
analysis shows that power spectra of galaxies and clusters are similar
in shape to the power spectrum of all matter, the main difference
being in the amplitude.  The power spectrum describes the square
of the amplitude of the density contrast, i.e. the amplitude of
density perturbations with respect to the mean density.  If we exclude
from the sample of all particles a population of approximately
constant density (void particles) and preserve all particles in
high-density regions, then the amplitudes of {\em absolute} density
fluctuations remain the same (as they are determined essentially by
particles in high-density regions), but the amplitudes of {\em
relative} fluctuations with respect to the mean density increase by a
factor, which is determined by the ratio of mean densities, i.e. by
the fraction of matter in the new density field with respect to the
previous one.  Our analysis has shown that actual differences in
amplitudes of power spectra of simulated galaxies with respect to the
power spectrum of all matter, are almost exactly equal to differences
calculated from the number of particles in respective samples,
i.e. the fraction of matter in high-density regions, $F_{c}$. This
fraction determines the biasing parameter of the sample of all
galaxies with respect to matter.

We have determined the fraction of matter in high-density regions
using numerical simulations of structure evolution for various
cosmological models.  Our analysis shows that the evolution is model
dependent: in models with high cosmological density voids are
evacuated more rapidly and less matter is left in high-density
regions.  A problem with numerical simulation of the void evacuation
is how to identify the present epoch in these simulations.  We have
done this using the calibration through the mean amplitude of density
fluctuations in a sphere of radius $r=8$~\Mpc, $\sigma_8$.  The
parameter $(\sigma_8)_{gal}$ can be determined directly from the
observed power spectrum of galaxies by integration, and it is related
to the corresponding parameter for matter, $(\sigma_8)_{m}$, through
an equation similar to the equation which relates amplitudes of power
spectra of matter and galaxies, with the fraction of matter in
galaxies (clustered matter).  Simulations yield another relation
between $(\sigma_8)_{m}$ and $F_{c}$, which allows us to determine
both unknown parameters.  We obtain $(\sigma_8)_{gal}=0.89 \pm 0.09$
for galaxies, $(\sigma_8)_m = 0.68 \pm 0.09$ for matter, and
$b_{gal}=1.3 \pm 0.13$ for the biasing factor of the clustered matter
(galaxies) relative to all matter.

\vskip0.5truecm
\noindent 
We thank A. Chernin, M. Gramann and D. Tucker for discussion, and
H. Andernach for suggestions to improve the text.  We thank the
referee, M. Vogeley, for constructive criticism. This work was
supported by the Estonian Science Foundation (grant 2625). JE and AS
were supported by the Deutsche Forschungsgemeinschaft in Potsdam; RC
was supported by grants NAG5-2759, AST9318185; AS was partially
supported by the Russian Foundation for Basic Research under Grant
96-02-17591.




\end{document}